\documentclass[final,5p,times,twocolumn]{elsarticle} 

\usepackage{lineno,hyperref}

\journal{Nuclear Instruments \& Methods in Physics Research, Section A}

\begin{document}

\begin{frontmatter}

\title{Novel Resistive-Plate WELL sampling element for (S)DHCAL}

\author{S. Bressler, P. Bhattacharya, A. Breskin, A.E.C. Coimbra, D. Shaked-Renous, A. Tesi}
\address{Weizmann Institute of Science, Rehovot, Israel}
\author{L. Moleri}
\address{Technion - Israel Institute of Technology, Haifa, Israel}
\author{M. Chefdeville, G. Vouters, J. Karyotakis, C. Drancourt}
\address{Univ. Grenoble Alpes, Univ. Savoie Mont Blanc, CNRS, IN2P3-LAPP, Annecy, France}
\author{M. Titov}
\address{CEA Saclay / Irfu, France}
\author{T. Geralis}
\address{NCSR Demokritos/INPP, Greece}

\begin{abstract}
Digital and Semi-Digital Hadronic Calorimeters (S)DHCAL were suggested for future Colliders as part of the particle-flow concept. Though studied mostly with RPC-based techniques, investigations have shown that Micro Pattern Gaseous Detector (MPGD)-based sampling elements could outperform in terms of average pad multiplicity or at higher rates. An attractive, industry-produced, robust, particle-tracking detector for large-area coverage, e.g. in (S)DHCAL, could be the novel single-stage Resistive Plate WELL (RPWELL). It is a single-sided  THick Gaseous Electron Multiplier (THGEM) coupled to the segmented readout electrode through a sheet of large bulk resistivity. 
We summarize here the preliminary test-beam results obtained with  6.5 mm thick (incl. electronics) {$48 \times 48\,\mathrm{cm^2}$}~RPWELL detectors. Two configurations are considered: a standalone RPWELL detector studied with 150 GeV muons and high-rate pions beams and a RPWELL sampling element investigated within a small-(S)DHCAL prototype consisting of 7 resistive Micro-MEsh Gaseous Structure (MICROMEGAS) sampling elements followed by 5 RPWELL ones. The sampling elements were equipped with a Semi-Digital readout electronics based on the MICROROC chip. 
\end{abstract}


\end{frontmatter}


\section{Introduction}

The particle-flow~\cite{Thomson:2009rp} is the leading concept towards reaching the challenging targeted jet energy resolution in future collider experiments ($\frac{\sigma_E}{E}=\frac{30\%}{\sqrt{E}}$~corresponding to $\frac{\sigma_E}{E}=3\%$~for 100 GeV jets). Particle-flow calorimeters~\cite{Thomson:2011zz, Sefkow:2015hna} are key ingredients in the design of experiments optimized for this concept. Having very high granularity, they allow separating the energy deposited by the individual constituents of the jets and measure the energy of each of them in the most adequate subsystem. Digital and Semi-Digital Hadronic Calorimeters ((S)DHCAL) are  attractive tools to achieve very high granularity while using cost-effective readout solutions. A typical (S)DHCAL consists of alternating layers of absorbers and sampling elements. Hadronic showers are mostly formed in the absorber, of which the material defines the total calorimeter's depth. The resulting signals are measured by sampling pad-readout elements (typically of  $1\,\mathrm{cm^2}$), defining the granularity.  
In (S)DHCAL, the measurement of the energy of individual particles relies on the approximate linear relation between the particle energy and the number of fired pads. Thus, the targeted jet-energy resolution calls for high detection efficiency at low average pad multiplicity. Detection elements based on the glass-RPC technology have been so far the most studied ones~\cite{Laktineh:2011zz, Repond:2012dwa,Buridon:2016ill}. Depending on the operation voltage, they can yield an average pad multiplicity of 1.5 - 2 at 90 - 95\% efficiency, in $1\,\mathrm{m^2}$~detectors~\cite{Repond:2012dwa,Buridon:2016ill}. Detection elements based on the MICROMEGAS have demonstrated superior properties: 98\% efficiency (at optimal operation voltage) with average pad multiplicity close to unity, in $1\,\mathrm{m^2}$~detectors~\cite{Adloff:2013wea,Adloff:2014qea} demonstrating also uniform response over the entire sensitive area. A detection efficiency of 95\% at similar average pad multiplicity was demonstrated with $16\times 16\,\mathrm{cm^2}$ resistive-MICROMEGAS prototypes, introduced to reduce the probability of discharges induced by highly ionizing particles~\cite{Chefdeville:2014rxa}. Elements based on double Gaseous Electron Multipliers (GEM), of a $1\,\mathrm{m^2}$ size, showed a multiplicity of $\sim$1.3 at 95\% efficiency~\cite{Yu:2011ik}.

\section{The Resistive Plate WELL}

\begin{figure}
\centering
  \includegraphics[width=0.45\textwidth]{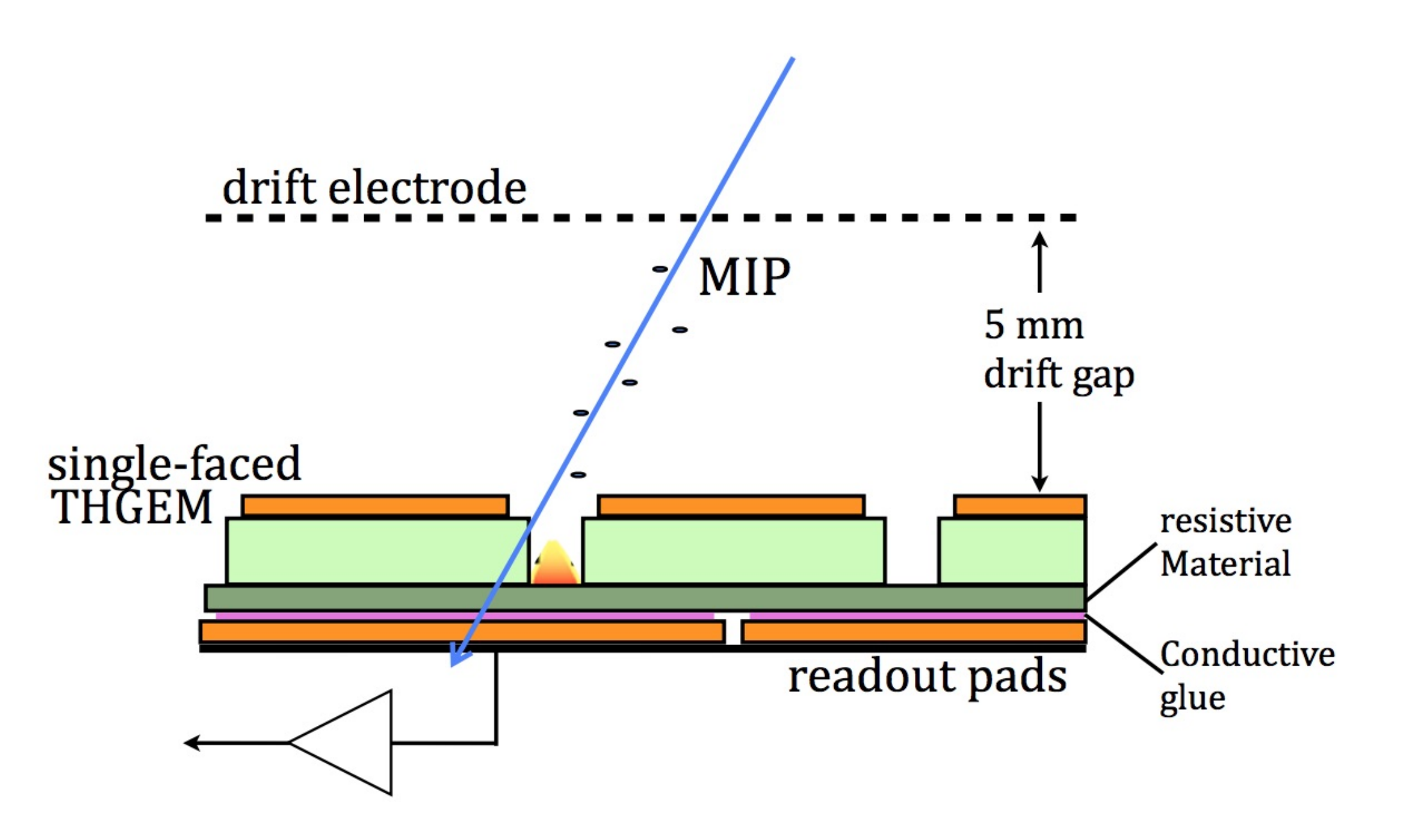}
  \caption{The Resistive Plate WELL (RPWELL) configuration with resistive anode and a (patterned) readout electrode. The WELL, a single-faced THGEM, is coupled to a copper anode via a resistive plate. Induced charges are collected from the anode before the avalanche charges traverse the resistive plate and evacuated through the anode.}
  \label{fig:RPWELL}
\end{figure}

\begin{figure}
\centering
  \includegraphics[width=0.40\textwidth]{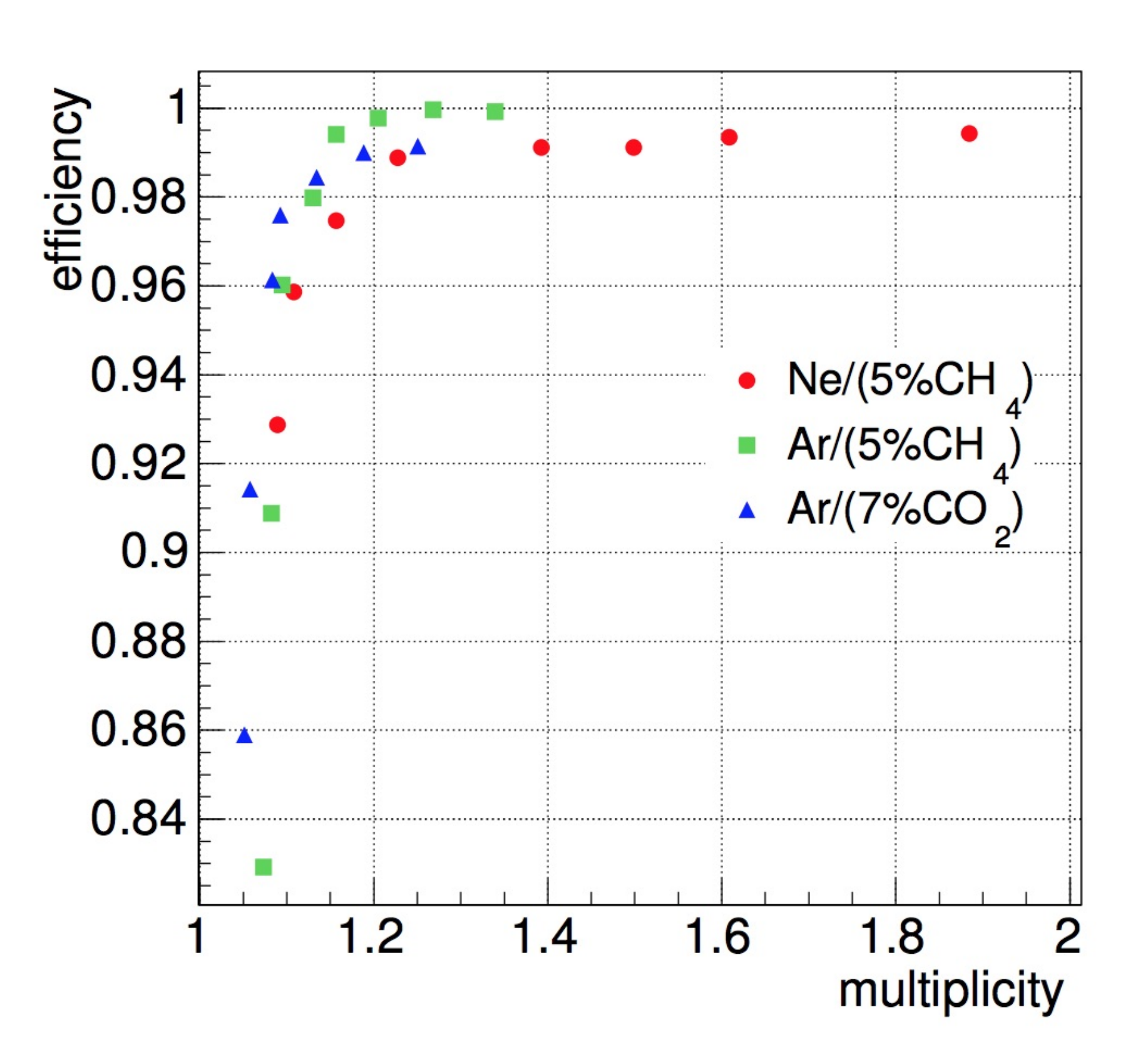}
  \caption{MIP detection efficiency as a function of average pad multiplicity measured with a $10 \time 10\,\mathrm{cm^2}$ RPWELL prototype in Ne- and Ar-based gas mixtures. Figure taken from~\cite{Moleri:2016bjv}.}
  \label{fig:Performance}
\end{figure}

The Resistive Plate WELL (RPWELL)~\cite{Rubin:2013jna} has followed a series of other Resistive THick Gaseous Electron Multiplier (THGEM)-based sampling elements developed over the past years at Weizmann Institute~\cite{Arazi:2013hdn,Bressler:2013cxa}. It is a robust, industrially mass-produced, single-stage particle-tracking gas-avalanche detector. With its discharge-free operation also in harsh radiation fields, large dynamic range, close-to-unity MIP detection efficiency and $\sim 200\,\mathrm{\mu m}$~RMS resolution~\cite{Moleri:2017qhi} – it becomes an attractive new candidate for particle tracking over large-area coverage. As a few-millimeter thin detector, it could become a candidate of choice as sampling element for (S)DHCAL.
The RPWELL (Fig.~\ref{fig:RPWELL}) is a single-sided THGEM electrode coupled to a segmented readout electrode through a thin sheet of large bulk resistivity ($10^8 - 10^{10}\,\mathrm{\Omega cm}$) material. The latter has the role of quenching large-size avalanches and preventing discharge development. Past laboratory and accelerator studies have been performed with moderate-size prototypes, with $1\,\mathrm{cm^2}$ square pads and SRS/APV25 readout electronics. They operated equally well in Ne- and Ar-based gas mixtures~\cite{Moleri:2016bjv} and in intense hadronic beams. The figure of merit is MIP detection efficiency $\geq 98\%$ at $\leq$ 1.2 pad multiplicity (Fig.~\ref{fig:Performance})~\cite{Moleri:2016bjv}. 

\begin{figure}
\centering
  \includegraphics[width=0.40\textwidth]{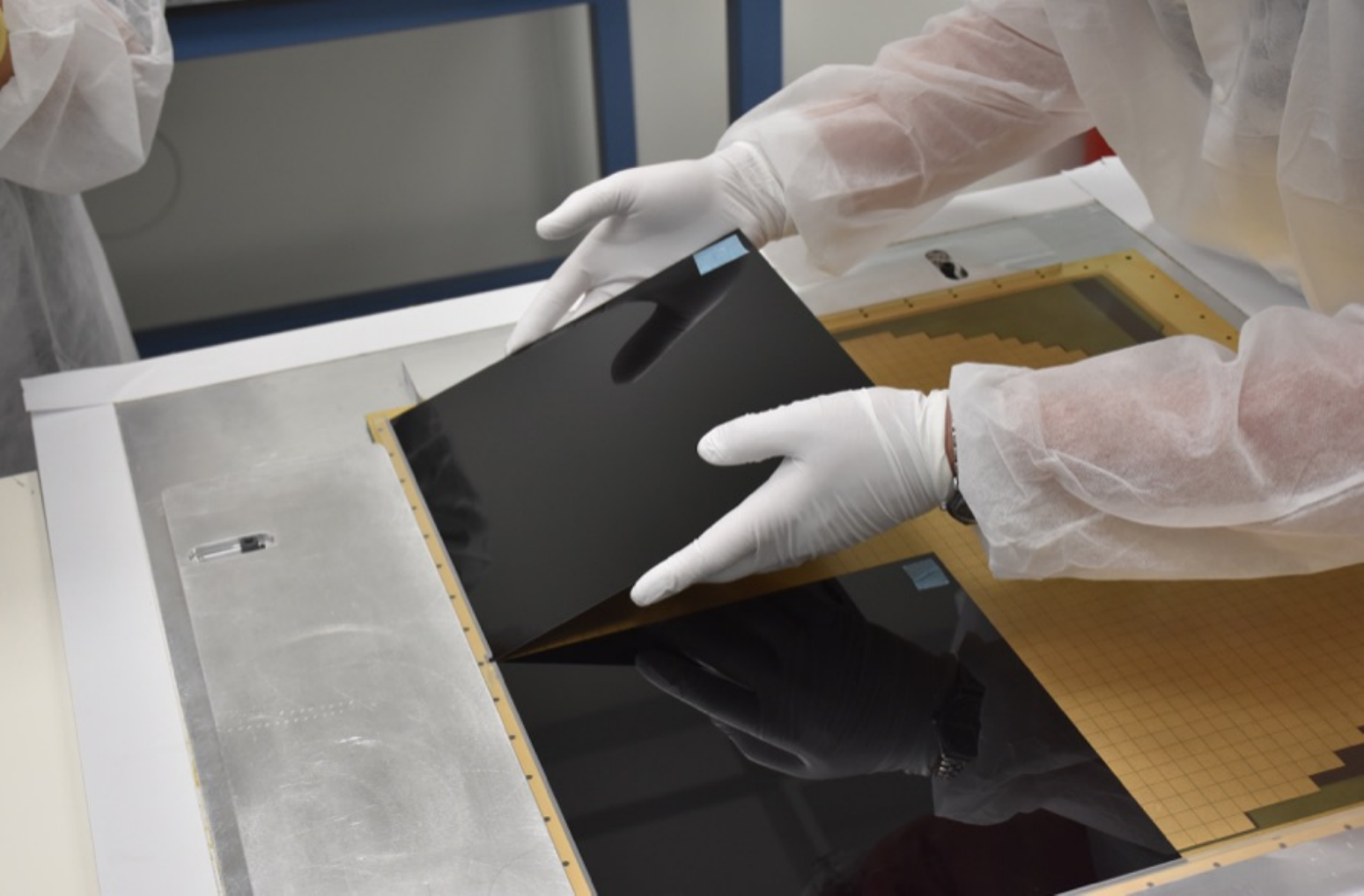}
  \caption{The gluing of the glass resistive plate to the pad-anode. Part of the assembly procedure.}
  \label{fig:assembly}
\end{figure}

Having in mind their application to (S)DHCAL, techniques were developed for producing large-area ($48 \times 48\,\mathrm{cm^2}$) 4.5 mm thick (excluding electronics) detectors, incorporating $10^{10}\,\mathrm{\Omega cm}$ silicate glass resistive plates (Fig.~\ref{fig:assembly}). Five such detectors were built and equipped with a pad-anode (defining a circular-shaped active area) embedding ILC-(S)DHCAL MICROROC chips~\cite{Adloff:2012zz} resulting in a total thickness of 6.5 mm. The RPWELL differed by their electrode quality, of which the thickness variation ranged from 5\% (best) to 25\% (worst), affecting significantly their stability and hence performance. 

\section{Performance in a standalone mode}
During August 2018, the first (S)DHCAL sampling element prototype built (with 25\% electrode thickness variation) has been investigated at CERN/SPS, in $\mathrm{Ar\/(7\%)CO_2}$, with muons and high-rate pions. Preliminary analysis results confirm that the performance of this prototype would be suitable for (S)DHCAL since a $\geq 95\%$ detection efficiency across most of the surface was achieved with a pad multiplicity of $\sim$1 in most events. The average pad multiplicity value was 1.7 due to a small number of events with tens of pads firing - probably indicating a discharge. Some efficiency variations as well as the small number of discharges are attributed to the large electrode-thickness variations (thus gain). 

\section{Performance within a small-(S)DHCAL prototype}
\begin{figure}
\centering
  \includegraphics[width=0.45\textwidth]{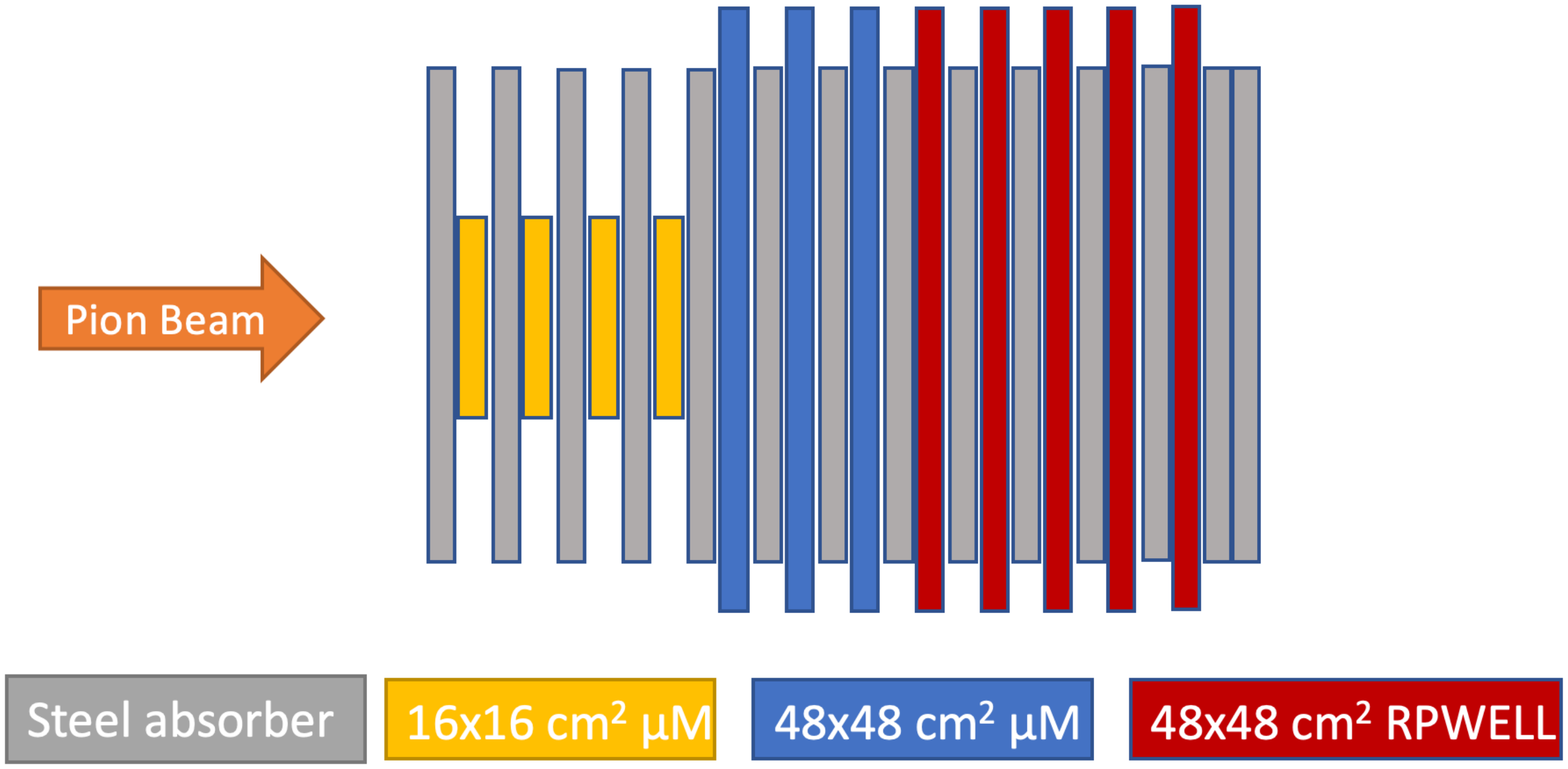}
  \caption{Small-(S)DHCAL prototype schematic description, consisting of four $16 \time 16\,\mathrm{cm^2}$ and three $48 \time 48\,\mathrm{cm^2}$  resistive MICROMEGAS followed by five $48 \time 48\,\mathrm{cm^2}$ RPWELL sampling elements. 2 cm thick stainless steel absorber planes were inter-layered between the sampling elements. }
  \label{fig:SDHCALproto}
\end{figure}

During November 2018, a small-(S)DHCAL prototype (Fig.~\ref{fig:SDHCALproto}) consisting of four $16 \times 16\,\mathrm{cm^2}$ bulk MICROMEGAs and three $48 \times 48\,\mathrm{cm^2}$ resistive MICROMEGAS sampling elements followed by five $48 \times 48\,\mathrm{cm^2}$ RPWELL ones has been investigate at CERN/PS using a low energy (2-6 GeV) pion beam. The 12 sampling elements were equipped with a semi digital readout electronics based on the MICROROC chip and read out with a single DAQ system. The RPWELLs with large thickness variations were excluded in some of the measurements. These were carried out with 8-layer (S)DHCAL consisting of 3 $16 \times 16\,\mathrm{cm^2}$ and 3 $48 \times 48\,\mathrm{cm^2}$ resistive MICROMEGAS sampling elements followed by 2 $48 \times 48\,\mathrm{cm^2}$ RPWELL ones.

\begin{figure}
\centering
  \includegraphics[width=0.45\textwidth]{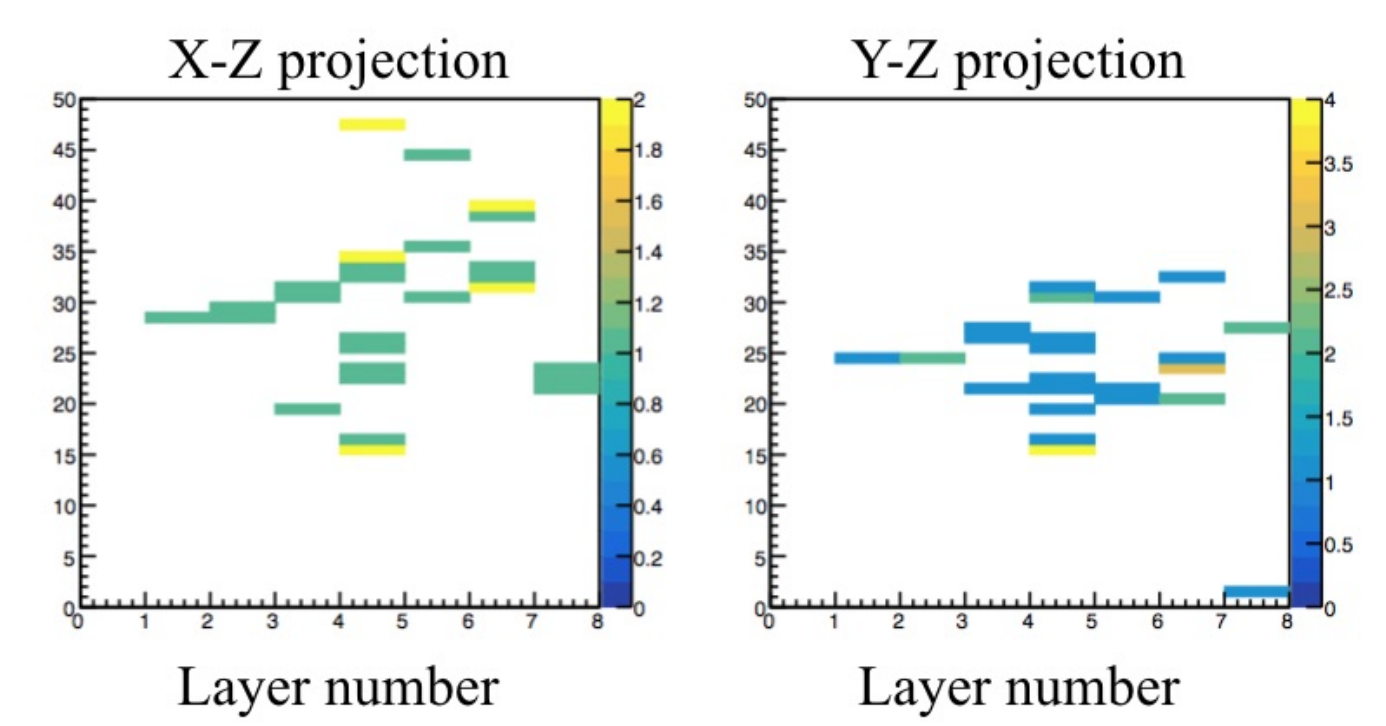}
  \caption{6 GeV pion shower recorded with the small-(S)DHCAL prototype of Fig.~\ref{fig:SDHCALproto}. Left: X-Z projection. Right: Y-Z projection. The Z-direction is defined as the beam direction.}
  \label{fig:shower}
\end{figure}

\begin{figure}
\centering
  \includegraphics[width=0.45\textwidth]{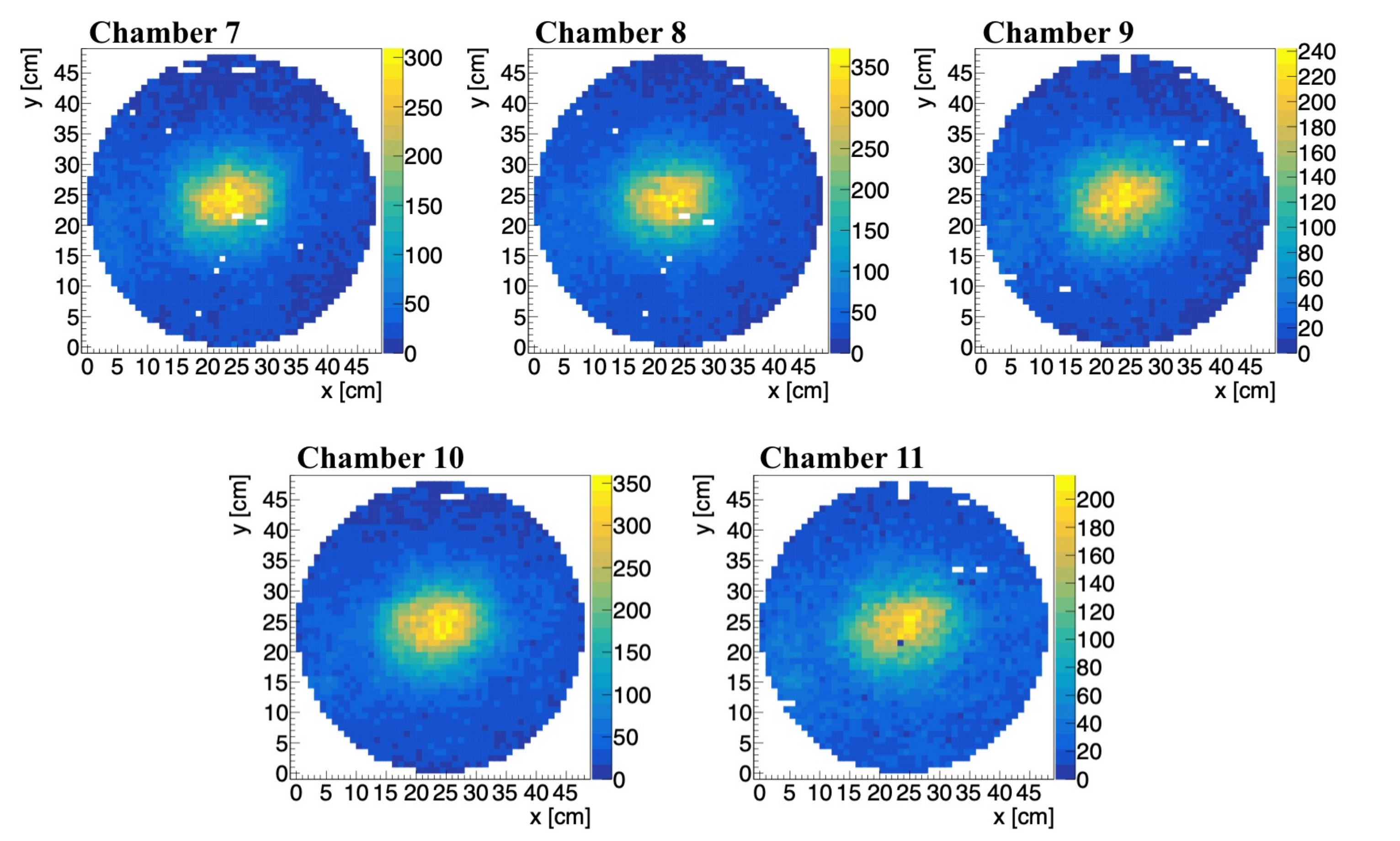}
  \caption{The beam profile recorded with the 5 RPWELL sampling elements. The label refers to the chamber location in the SDHCAL prototype (Fig.~\ref{fig:SDHCALproto})}
  \label{fig:beamProfile}
\end{figure}

The hits associated with a single shower are grouped based on time selection. 
A pion shower profile recorded in all the sampling elements of the 8-layer prototype and a beam profile in the 5 RPWELL detectors are shown in Fig.~\ref{fig:shower} and~\ref{fig:beamProfile} respectively. 

\begin{figure}
\centering
  \includegraphics[width=0.40\textwidth]{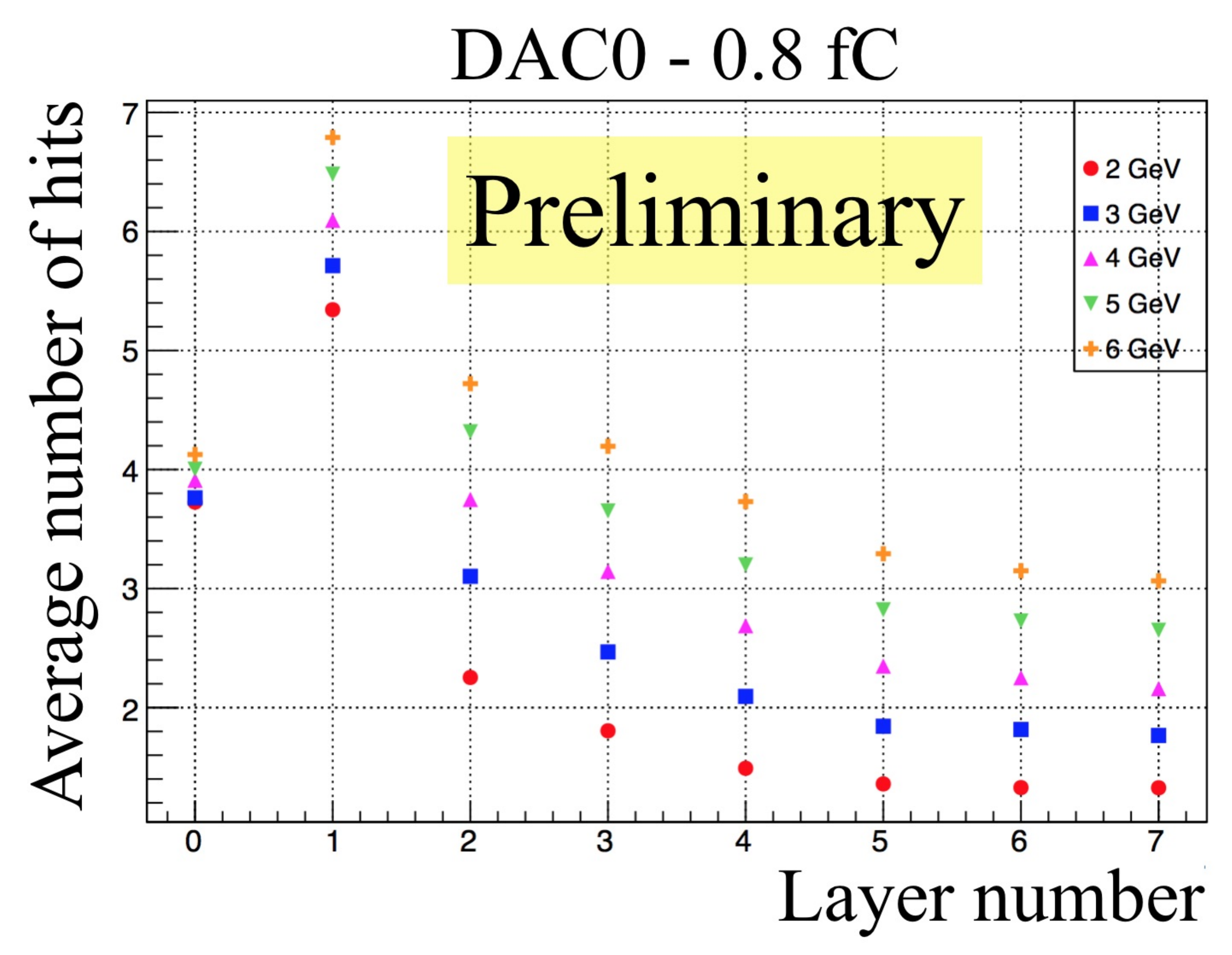}
  \caption{The average number of RPWELL pads fired (charge above 0.8 fC) as a function of the shower depth (layer number). The RPWELL layer number is relative to the shower starting layer.}
  \label{fig:hitsPerLayer}
\end{figure}

\begin{figure}
\centering
  \includegraphics[width=0.40\textwidth]{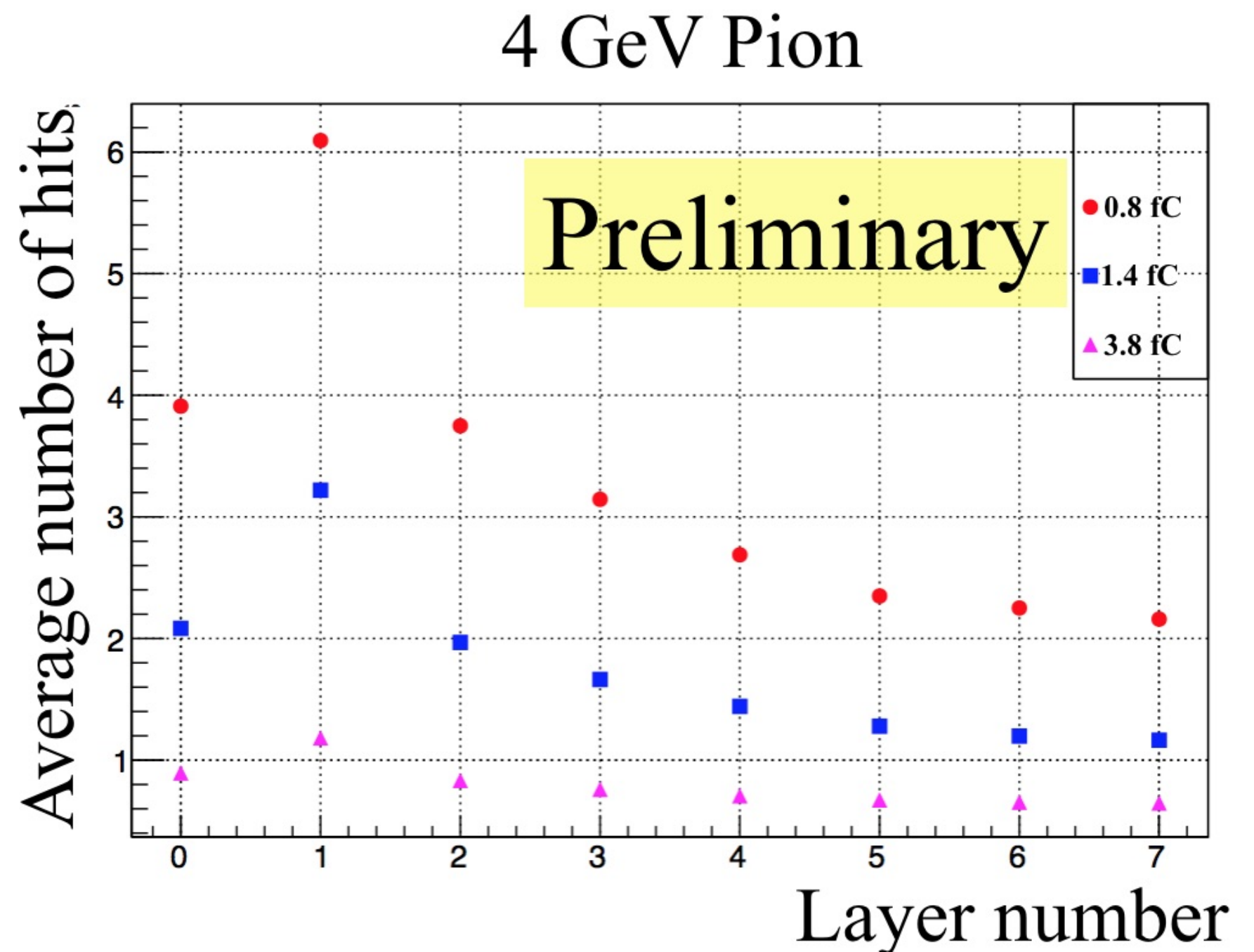}
  \caption{The average number of RPWELL pads fired by 4 GeV pions as a function of the shower depth (layer number). The RPWELL layer number is relative to the shower starting layer. Results are shown for three threshold levels: 0.8, 1.4 and 3.8 fC.}
  \label{fig:4GeVPionHitsPerLayer}
\end{figure}

\begin{figure}
\centering
  \includegraphics[width=0.40\textwidth]{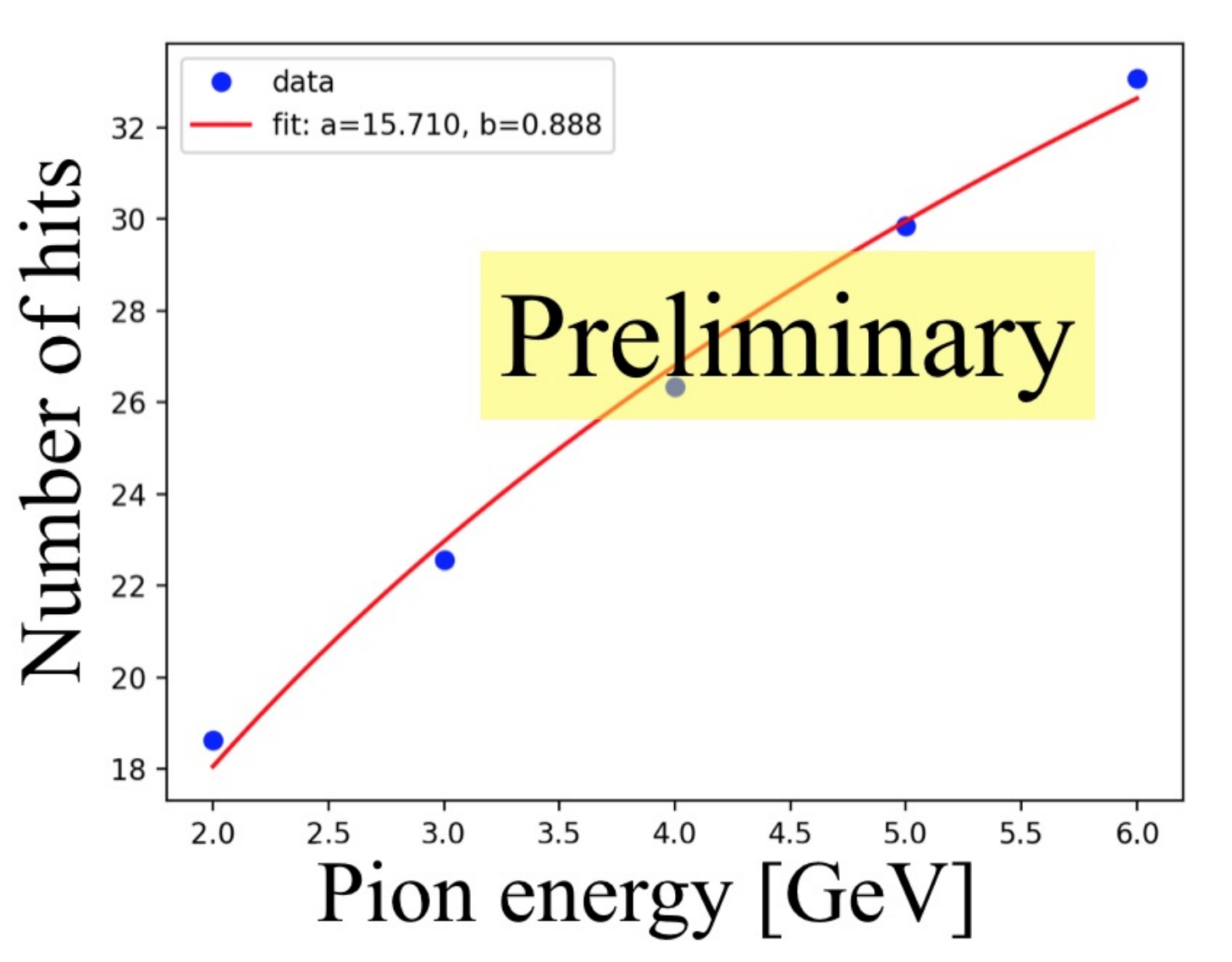}
  \caption{The expected average total number of hits recorded by a full RPWELL-based small-(S)DHCAL as a function of the incoming pion energy.}
  \label{fig:hitsVsE}
\end{figure}

For each recorded shower, the shower origin was defined as the layer in which at least three pads fired. This layer was  denoted by \textit{Layer 0} and it served as a reference to define the depth (layer number) within the shower of the RPWELL sampling elements. The average number of RPWELL pads fired as a  function of the shower depth  is shown in Fig.~\ref{fig:hitsPerLayer} for different incoming-pion energies, with the lowest threshold level (0.8 fC) applied. The average number of pads fired as a function of the shower depth is shown in Fig.~\ref{fig:4GeVPionHitsPerLayer} for 4 GeV pions, at the three threshold levels. A preliminary evaluation of the total number of hits that would be recorded by a small-(S)DHCAL prototype consisiting of only RPWELL-based sampling elements as a function of the incoming pion energy is shown in Fig.~\ref{fig:hitsVsE}. For each pion energy, the total number of hits is estimated as the sum of the average number of hits recorded as a function of the shower depth. Some of the observed non-linearity could be attributed to the leakage from the small-(S)DHCAL assembly. 

\section{Summary and discussion}
First studies of RPWELL-based sampling elements for (S)DHCAL have been carried out in a standalone mode and within a small-(S)DHCAL prototype (incorporating also MICROMEGAS-based sampling elements). 
More stringent QA/QC tests will have to be applied in the future to ensure good control of the thickness of the WELL electrode to a level better than 5\%. 
The $48 \times 48\,\mathrm{cm^2}$ RPWELL was operated under 150 GeV muon and high rate pion beams. Up to some instabilities attributed to the thickness variations, detection efficiency greater than 95\% and an average pad multiplicity close to "1" were recorded. In the small-(S)DHCAL prototype, low energy pion showers were recorded, and the response of the RPWELL is consistent with the shower depth. Based on the data collected, the estimation of the expected pion energy resolution in a full RPWELL-based (S)DHCAL is ongoing.

\section{Acknowledgments}
We thank Prof. Wang Yi from Tsinghua university in China for kindly providing us with the silicate glass tiles.
This research was supported in part by the I-CORE Program of the Planning and Budgeting Committee,
the Nella and Leon Benoziyo Center for High Energy Physics at the Weizmann Institute of Science, the common fund of the RD51 collaboration at CERN (the Sampling Calorimetry with Resistive Anode MPGDs, SCREAM,
 project), Grant No 712482 from the Israeli Science Foundation (ISF) and Grant No 5029538 from the Structural Funds, ERDF and ESF, Greece.

\section*{References}

\bibliography{mybibfile}

\end{document}